\documentclass[10pt,conference]{IEEEtran}

\usepackage{amsmath,url}
\usepackage{epsfig,amssymb,amsbsy,verbatim,array}
\usepackage{pstricks,psfrag,theorem,cite,enumerate}%

\let\intern=\iftrue

\def\figref#1{Fig.\,\ref{#1}}%
\def\E{\mathbb{E}}
\def\P{\mathbb{P}}
\def\R{\mathbb{R}}
\def\Z{\mathbb{Z}}

\def\L{\mathbb{L}}
\def\ie{{\em i.e.}}

\def\d{\textnormal{d}}

\def\rg{r_{\mathrm G}}
\def\bNin{\bar N^{\mathrm{in}}}
\def\bNout{\bar N^{\mathrm{out}}}
\def\Nin{N^{\mathrm{in}}}
\def\Nout{N^{\mathrm{out}}}

\DeclareMathAlphabet{\mathsfsl}{OT1}{cmss}{m}{sl}

\theoremstyle{plain}
\newtheorem{theorem}{Theorem}

\newtheorem{proposition}[theorem]{Proposition}
\newtheorem{fact}[theorem]{Fact}
\newtheorem{conjecture}[theorem]{Conjecture}

\renewcommand{\baselinestretch}{1.00}
\begin{document}
\title{The Secrecy Graph and Some of its Properties}
\author{Martin Haenggi\\
Department~of
Electrical~Engineering\\ University of Notre Dame\\ Notre Dame, IN
46556, USA\\ E-mail: {\tt mhaenggi@nd.edu} }
\maketitle

\begin{abstract}
A new random geometric graph model, the so-called {\em secrecy graph}, is introduced
and studied.  The graph represents a wireless network and includes only edges
over which secure communication in the presence of eavesdroppers is possible. 
The underlying point process models considered are lattices and Poisson
point processes. In the lattice case, analogies to standard bond and site percolation
can be exploited to determine percolation thresholds. In the Poisson
case, the node degrees are determined and percolation is studied using
analytical bounds and simulations. It turns out that a small density of eavesdroppers
already has a drastic impact on the connectivity of the secrecy graph.
\end{abstract}

\section{Introduction}
There has been growing interest in information-theoretic secrecy.
To study the impact of the secrecy constraint on the connectivity
of ad hoc networks,
we introduce a new type of random geometric graph, the so-called
{\em secrecy graph}, that represents the network or communication graph
including only links over which secure communication is possible.
We assume that a transmitter can choose the rate very close to the
capacity of the channel to the intended receiver, so that any eavesdropper
further away than the receiver cannot intercept the message.
This translates into a simple geometric constraint for secrecy
which is reflected in the secrecy graph.
In this initial investigation, we study some of the properties of the secrecy graph.

\section{The Secrecy Graph}
Let $\hat G=(\phi,\hat E)$ be a geometric graph in $\R^d$, where
$\phi=\{x_i\}\subset\R^d$ represents the locations of the nodes, also referred to
as the ``good guys". We can think of this graph as the unconstrained network graph that
includes all possible edges over the good guys could communicate if there were no
secrecy constraints.

Take another set of points
 $\psi=\{y_i\}\subset\R^d$ representing the locations of the eavesdroppers or
 ``bad guys".  These are assumed to be known to the good guys.
 
Let $D_x(r)$ be the (closed) $d$-dimensional ball of radius $r$
centered at $x$, and let
$\delta(x,y)=\|x-y\|$ be a distance metric, typically Euclidean distance.
Further, let $\phi(A)$ and $\psi(A)$ 
denote the number of points of $\phi$ or $\psi$ falling in $A\subset\R^d$.

Based on $\hat G$, we define the following secrecy graphs (SGs):

{\em The directed secrecy graph:} $\vec G=(\phi,\vec E)$. Replace all edges
in $\hat E$ by two directional edges. Then remove all edges
$\overrightarrow{x_ix_j}$ for which $\psi\left(D_{x_i}(\delta(x_i,x_j))\right)>0$, \ie,
there is at least one eavesdropper in the ball.

From this directed graph, two undirected graphs are derived:

{\em The basic secrecy graph:} $G=(\phi,E)$, where the (undirected) edge set $E$
is
\[ E\triangleq \{ x_ix_j\,: \: \overrightarrow{x_ix_j}\in\vec E
\text{ and }\overrightarrow{x_jx_i}\in\vec E \} \,. \]

{\em The enhanced secrecy graph:} $G'=(\phi,E')$, where
\[ E'\triangleq \{ x_ix_j\,: \: \overrightarrow{x_ix_j}\in\vec E
\text{ or }\overrightarrow{x_jx_i}\in\vec E \} \,. \]

\begin{figure}
\parbox[c]{.3\columnwidth}{%
\centering
\epsfig{file=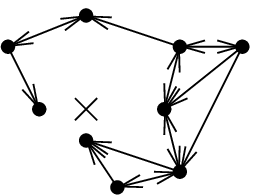,width=.3\columnwidth}
\scriptsize (a) Directed SG $\vec G$}\hfill
\parbox[c]{.3\columnwidth}{%
\centering\epsfig{file=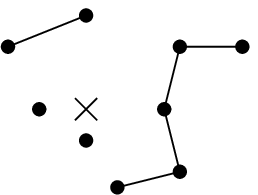,width=.3\columnwidth}
\scriptsize(b) Basic SG $G$}\hfill
\parbox[c]{.3\columnwidth}{%
\centering\epsfig{file=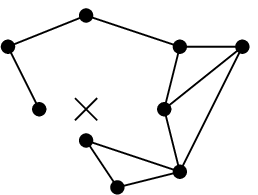,width=.3\columnwidth}
\scriptsize(c) Enhanced SG $G'$}
\caption{Example for secrecy graphs. The dots are
the good guys, the cross the eavesdropper. The underlying graph is assumed to be
 fully connected so that the secrecy graphs include all edges along which secure
 communication is possible.}
\label{fig:secrecy}
\end{figure}

Clearly, $E\subset E'$. The difference is that edges in $E$ permit
secure bidirectional communication while edges in $E'$ only allow
secure communication in one direction. However, this one-way link
may be used to transmit a one-time pad so that the other node can
reply secretly. In doing so,  the link capacity would drop from $1/2$ in
each direction to $1/3$.
\figref{fig:secrecy} shows an example in $\R^2$ for the three types of secrecy graphs.
based on the same underlying fully connected graph $\hat G$.

One way to assess the impact of the
secrecy requirement is to determine the {\em secrecy ratios}
\begin{equation}
\eta=\frac{|E|}{|\hat E|}=\frac{\bar N}{\bar{\hat N}}\,;\qquad 
   \eta'=\frac{|E'|}{|\hat E|}=\frac{\bar N'}{\bar{\hat N}}\,,
\label{eta}
\end{equation}
where $\bar N$, $\bar N'\geqslant \bar N$, and $\bar{\hat N}$ are the average node
degrees of the respective graphs.
For $\eta\approx 1$, the impact of the secrecy requirement is negligible while for small
$\eta$ it severely prunes the graph.

For the directed graph,
the mean in- and out-degrees are equal, so we define
$\vec N\triangleq \bNout=\bNin=|\vec E |/|\phi|$.
Since the edge sets of $G$ and $G'$ are a partition of the edge set
of the undirected multigraph containing all edges in $\vec G$,
the following holds:
\begin{fact}
The mean degrees are related by
\begin{equation}
  \bar N+\bar N'=2\bNin=2\bNout\,.
  \label{degree_relation}
\end{equation}
Furthermore, the degrees of all nodes $x\in \phi$ are bounded by
\begin{multline}
 N_x\leqslant \min\{\Nin_x,\Nout_x\} \leqslant \max\{\Nin_x,\Nout_x\} \\
   \leqslant N'_x \leqslant \Nin_x+\Nout_x\,.
\end{multline}
\end{fact}
In the example in \figref{fig:secrecy}, \eqref{degree_relation} yields
$10/9+22/9=32/9$.

These graphs become interesting if the locations of the vertices are
stochastic point processes. We will use $\Phi$ and $\Psi$ as the
corresponding random variables.
 
Our goal is to study the properties of the resulting random geometric graphs,
including degree distributions and percolation thresholds.
We will consider two cases, lattices and Poisson
point processes.

\subsection{Lattice model}
Let the underlying graph be the standard square lattice in $\Z^2$, \ie,
$\hat G=\L^2$, where edge exists between all pairs of points with Euclidean distance $1$.
Let $\Psi$ be obtained from random independent thinning
of a regular point set where each point exists with probability $p$, independently
of all others.
Let the corresponding secrecy graphs be denoted as
$\vec G_p$, $G_p$, and $G'_p$. Let $\theta(p)$ be the probability that
the component containing the origin is infinite\footnote{In the directed case,
to be precise, we consider oriented percolation and let $\theta(p)$
be the probability that the out-component
is finite, \ie, that there are directed paths
from the origin to an infinite number of nodes. Alternatively (or in addition) we could
consider the in-component of the origin.}. Then
the percolation threshold is defined as
\begin{equation}
  p_c=\inf \{p\,:\: \theta(p)=0\}\,.
\end{equation}

\subsection{Poisson model}
Let the underlying graph
be Gilbert's disk graph $\hat G_r$ \cite{Gil}, where
$\Phi$ is a Poisson point process (PPP) of intensity $1$ in $\R^2$ and
two vertices are connected if their distance is at most $r$.
$\Psi$ is another, independent, PPP of intensity $\lambda$. 
Denote the secrecy graphs by $\vec G_{\lambda,r}$, $G_{\lambda,r}$,
and $G'_{\lambda,r}$. With $\theta(\lambda,r)$ being the probability that
the component containing the origin (or any arbitrary fixed node) is
infinite, the percolation threshold radius for $\hat G_r$ is
\begin{equation}
  \rg\triangleq\sup \{r\,:\: \theta(0,r)=0\} \approx 1.198\,,
\end{equation}
where the subscript G indicates that this is Gilbert's critical radius
which is not known exactly but the bounds $1.1979 < r_c < 1.1988$
were established with $99.99\%$ confidence in \cite{net:Balister05}.
For radii larger than $\rg$, we define
\begin{equation}
  \lambda_c(r)\triangleq\inf \{\lambda\,:\: \theta(\lambda,r)=0\}\,,\quad r>\rg.
\end{equation}

For the analyses, we assume 
that there is a node in $\Phi$ at the origin. This does not affect the distributional
properties of the PPP.

The parameter $r$ indicating the maximum range of transmission can be related
to standard communication parameters as follows:
Assume a standard path loss law with attenuation
exponent $\alpha$, a transmit power $P$, a noise floor $W$, and
a minimum SNR $\Theta$ for reliable communication. Then
\begin{equation}
  r=\left(\frac P{\Theta W}\right)^{1/\alpha}\,.
\end{equation}

\section{Properties of the Lattice Secrecy Graphs}
Consider first the graph $G'_p$ for the case where a bad guy is placed (with probability $p$)
in the middle of each edge in $\L^2$, \ie, 
\begin{equation}
  \Psi=\{ x\in (\mathbb{Z}^2+(1/2,0))\cup (\mathbb{Z}^2+(0,1/2))\,:\: U_x<p \}\,,
  \label{site}
\end{equation}
where $U_x$ is iid uniformly distributed in $[0,1]$.

This case is analogous to bond percolation on the two-dimensional
square lattice \cite{net:Grimmett99}, so:
\begin{fact}
  The percolation threshold for $G'_p$ is $p_c=1/2$.
\end{fact}
For $p=1/2$, the density of bad guys is $1$, the same as the density of good guys.
So the density of bad guys only needs to be a factor $1-\epsilon$ smaller than
the density of good guys, and the graph still percolates, for any $\epsilon>0$.
Close to that threshold $\eta'=1/2$.

If we put the bad guys on the lattice points themselves (with probability $p$),
the situation is analogous to site percolation on $\L^2$, for which the
critical probability is unknown but estimated to be around 0.59, so
$p_c\approx 0.41$. In this case the maximum density of
bad guys is only 0.41. 

So it seems having the eavesdroppers at the locations indicated
in \eqref{site} causes the least ``damage" to the connectivity of
the secrecy graph among all regular $\Psi$.

\section{Properties of the Poisson Secrecy Graphs}

\subsection{Isolation probabilities for $r=\infty$}
\begin{fact}
 The probability that the origin $o$ cannot talk to anyone in $\vec G_{\lambda,\infty}$ (out-isolation) is
 \begin{equation}
  \P[\Nout=0]=\frac{\lambda}{\lambda+1}\,.
  \label{out_isol}
 \end{equation}
 In $G_{\lambda,\infty}$:
 \begin{equation}
  \P[N=0]=\frac{c\lambda}{c\lambda+1}
  \label{basic_isol}
 \end{equation}
 where $c=\frac43+\frac{\sqrt3}{2\pi}= 1.609$.
 \end{fact}
 For the directed graph, this is simply the probability that the nearest neighbor in the
 combined process $\Phi\cup \Psi$ (of density $1+\lambda$) is
 an eavesdropper. In the basic graph, let $x$ denote the origin's nearest 
 neighbor in $\Phi$ and let $R=\|x\|$. For $N>0$, we need $D_o(R)\cup D_x(R)$ to be free
of eavesdroppers.
The area of the two intersecting disks is $c \pi R^2$ for
$c=4/3+\sqrt{3}/(2\pi)$, and the probability density (pdf) of 
$R$ is $p_R(r)=2\pi r\exp(-\pi r^2)$  (Rayleigh with mean 1/2) \cite{net:Haenggi03it}.
 So
\begin{equation}
 \P[N>0]=\E_R[\exp(-\lambda c\pi R^2)]=\frac{1}{c\lambda+1}\,.
\end{equation}
The probability of in-isolation $\P[\Nin=0]$ is smaller than $\P[\Nout=0]$, since
for each node $x\in\Phi\backslash\{o\}$, there must be at least one bad guy
in $D_x(\|x\|)$. Clearly, this probability is smaller compared to \eqref{out_isol}
where only the nearest eavesdropper matters. On the other hand,
the in-degree is less likely than the out-degree to be large since a significantly
larger area needs to be free of bad guys.
The isolation probability $\P[N'=0]$ is the smallest of all isolation probabilities.

\subsection{Degree distributions}
\begin{proposition}
 The out-degree of $o$ in $\vec G_{\lambda,\infty}$ is geometric with mean $1/\lambda$.
 \label{prop:degree}
\end{proposition}
\begin{IEEEproof}
Consider the sequence of nearest neighbors of $o$ in the combined
process $\Phi\cup\Psi$. $\Nout=n$ if the closest $n$ are in $\Phi$
and the $(n+1)$-st is in $\Psi$. Since these are independent events,
\begin{equation}
 \P[\Nout=n]=\frac{\lambda}{1+\lambda}\left(\frac 1{1+\lambda}\right)^n\,.
 \label{geom}
\end{equation}
\end{IEEEproof}
Alternatively, the distribution can be obtained as follows:
Let $R$ be the distance to the closest {\em bad} guy, \ie,
$R=\min_{y\in\Psi} \|y\|$.
We have
\begin{equation}
\P[\Nout=n]=\E_R \left[\exp(-\pi R^2)\frac{(\pi R^2)^n}{n!}\right]\,,
\end{equation}
where the pdf of $R$ is $p_R(r)=2\pi r\lambda\exp(-\pi \lambda r^2)$.

For general $r$, we have:
\begin{proposition}
 The out-degree distribution of $\vec G_{\lambda,r}$ is
 \begin{equation}
\P[\Nout=n] =\frac{\lambda\left(1-\frac{\Gamma(n,a)}{\Gamma(n)}\right)+\exp(-a) \frac{a^n}{n!}}{(\lambda+1)^{n+1}}\,,
\label{out_dist}
\end{equation}
where $a=\pi r^2(\lambda+1)$, and $\Gamma(\cdot,\cdot)$ is the {\em upper} incomplete
gamma function.
The probability of out-isolation is
\begin{equation}
\P[\Nout=0]=\frac{\exp(-\pi r^2 (\lambda+1))+\lambda}{1+\lambda}\,,
\label{void}
\end{equation}
and the mean out- and in-degrees are
\begin{equation}
 \E \Nout=\E\Nin=\frac 1\lambda (1-\exp(-\lambda\pi r^2))\,.
 \label{mean_deg_dir}
\end{equation}
The mean degree of the basic graph $G_{\lambda,r}$ is
\begin{equation}
  \E N=\frac{1}{c\lambda}(1-\exp(-c\lambda\pi r^2))\,,
  \label{mean_deg_basic}
\end{equation}
where $c=\frac 43+\frac{\sqrt 3}{2\pi}$.
\end{proposition}
\begin{IEEEproof}
If there is no bad guy
inside $D_x(r)$ --- which is the case with probability $P_0=\exp(-\lambda\pi r^2)$ ---
then the number is simply Poisson with mean $\pi  r^2$.
If there is a bad guy at distance $R$, the number is Poisson with mean 
$\pi R^2$. So we have
\begin{align*}
  \P[\Nout=n]= &P_0\exp(-\pi r^2)\frac{(\pi r^2)^n}{n!} \\
    &+(1-P_0)\E_{R< r}
    \left[\exp(-\pi R^2)\frac{(\pi R^2)^n}{n!}\right]\,,
\end{align*}
which, after some manipulations, yields \eqref{out_dist}. The mean is obtained
more directly using Campbell's theorem \cite{net:Stoyan95} which
says that for
stationary point processes with intensity $\mu$ in $\R^2$ and
non-negative measurable $f$,
\[  \E \left[ \sum_{x\in\Phi} f(x)\right]=\mu \int_{\R^2} f(x) \d x \,. \]
Applied to the mean out-degree ($\mu=1$) we obtain
\begin{align}
\E\Nout&=\sum_{x\in\Phi}\P(\overrightarrow{ox}\in \vec E)
 = \E \sum_{\substack{x\in\Phi\backslash \{o\} \\ \|x\|<r}} \exp(-\lambda\pi \|x\|^2)\nonumber\\
  &=
     2\pi\int_0^r x\exp(-\lambda\pi x^2)\d x\,.
\end{align}
By replacing the area $\pi r^2$ by $c\pi r^2$ (area of two overlapping disks of 
radius $r$ and distance $r$), the same calculation yields \eqref{mean_deg_basic}.
\end{IEEEproof}

\begin{figure}
\centering
\epsfig{file=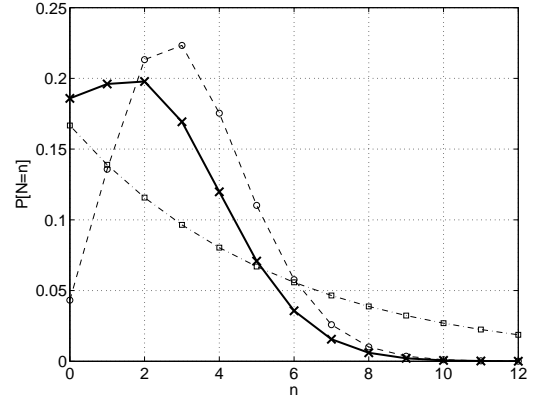,width=7cm}
\caption{Distribution of the node degree with and without power constraints.
The solid bold curve shows the distribution \eqref{out_dist} for $r=1$ and $\lambda=1/5$.
The dashed curve is the Poisson distribution with mean $\pi$ (which results when $r=1$, $\lambda=0$),
and the dash-dotted curve is the geometric distribution with mean $5$ (which results when
$\lambda=1/5$, $r\rightarrow\infty$).}
\label{fig:deg_dist}
\end{figure}

\begin{figure}
\centering
\epsfig{file=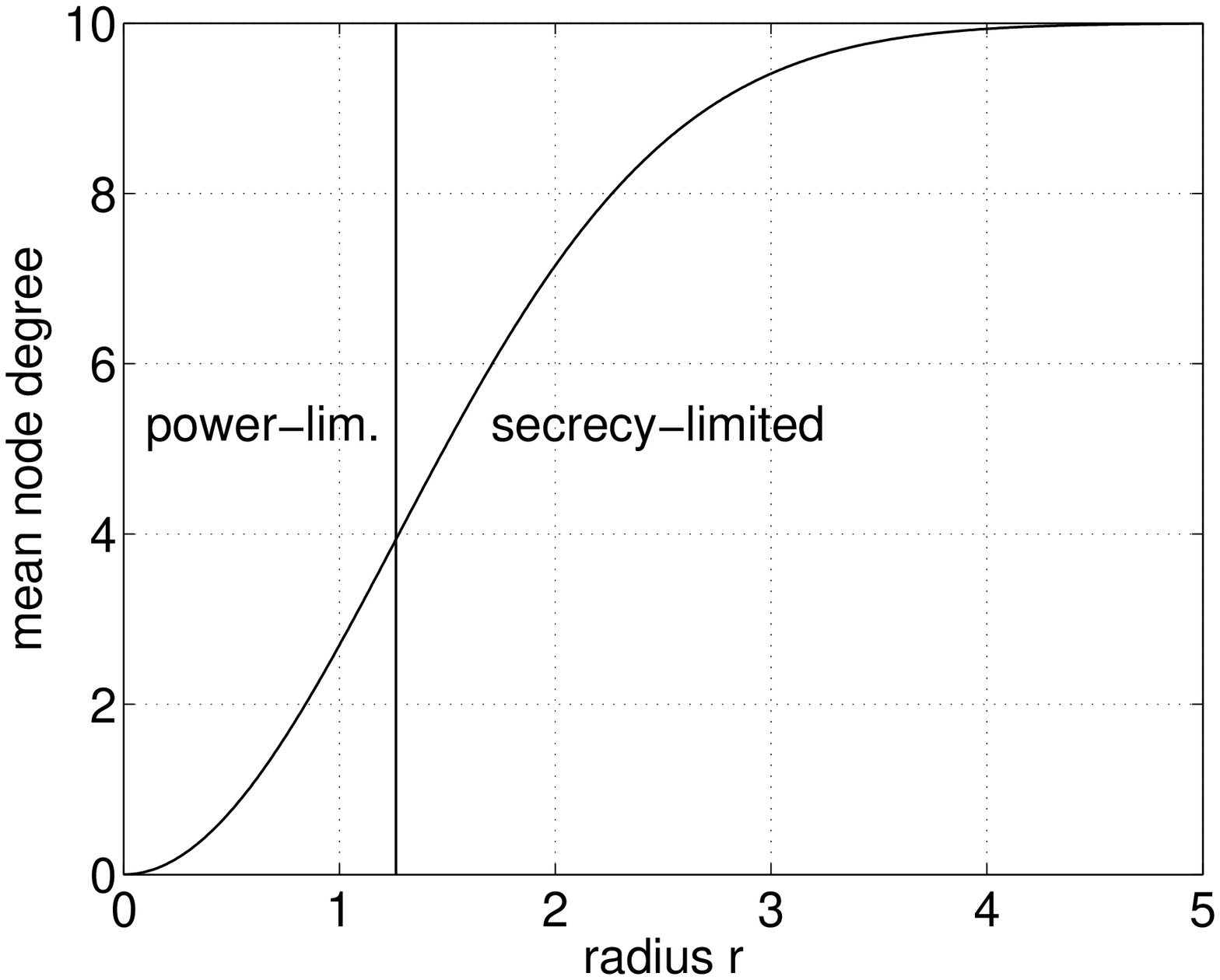,width=.48\columnwidth}\hfill
\epsfig{file=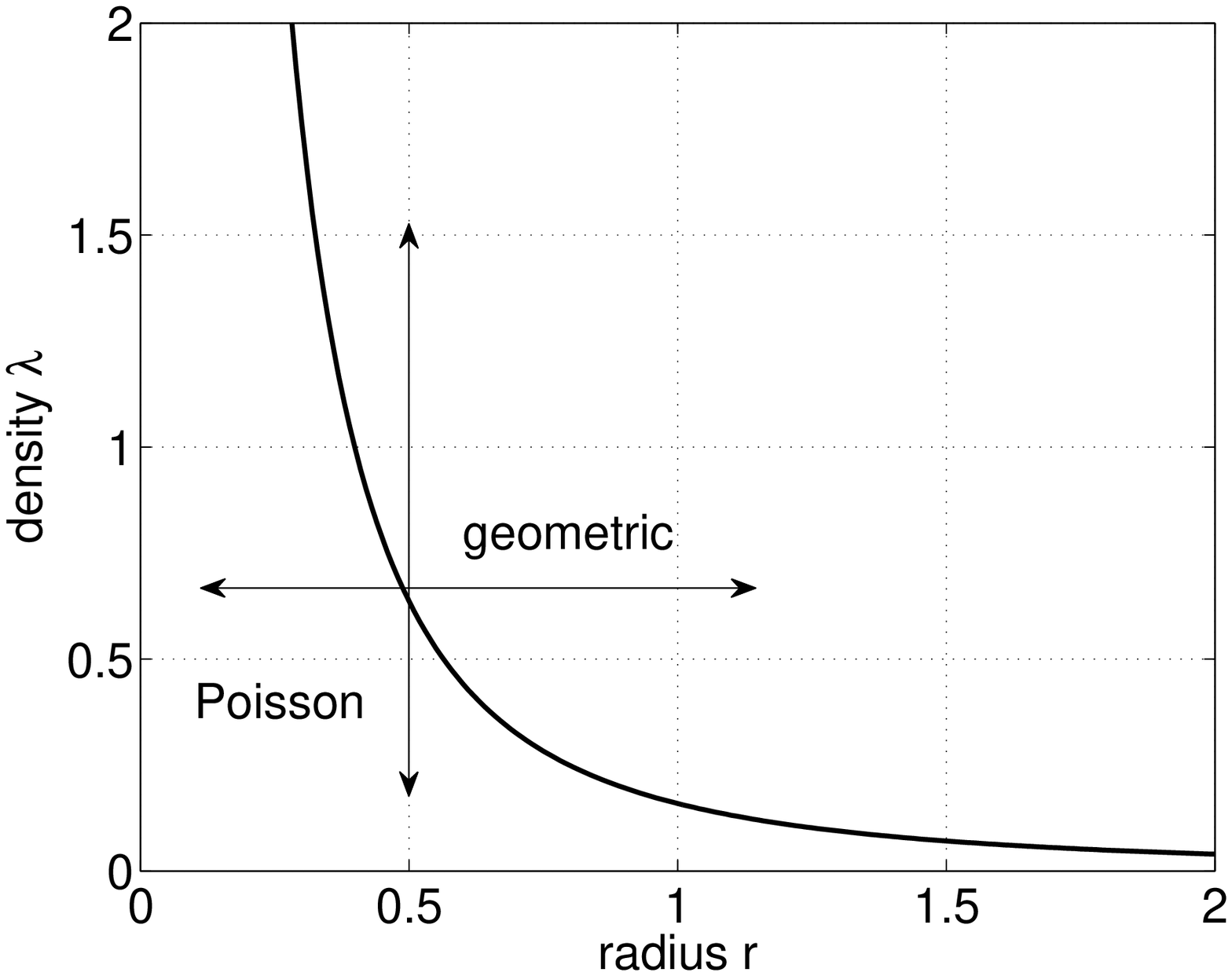,width=0.48\columnwidth}
\caption{Left: Mean out-degree of $\vec G_{1/10,r}$. The vertical line goes through
the inflection point and indicates the boundary between the power-limited and the
secrecy-limited regime. At the inflection point, $r=r_T=(2\pi\lambda)^{-1/2}=\sqrt{5/\pi}\approx 1.26$.
\hfill
Right: The curve $\lambda=(2\pi r^2)^{-1}$ bordering the power-limited and the secrecy-limited
regimes. As $r\rightarrow\infty$ or $\lambda\rightarrow\infty$, the network becomes
secrecy-limited and the degree distribution is geometric. As $r\rightarrow 0$ or $\lambda\rightarrow 0$,
the network is power-limited and the degree distribution is Poisson.}
\label{fig:mean_degree}
\end{figure}

The probability of isolation can directly be obtained from considering the two
possibilities for isolation: Either there is no node at all within distance $r$ or there
is one node (or more) within $r$ and the nearest one is bad. So, using $a$ as in
the proposition, $\P[\Nout=0]=\exp(-a)+(1-\exp(-a))\lambda/(1+\lambda)$.
Since $\Phi$ has intensity $1$, the isolation probability equals the density
of isolated nodes.

As $\lambda\rightarrow 0$, we obtain the Poisson
isolation probability and for $r\rightarrow\infty$ we get the
geometric isolation probability \eqref{geom}. 
Also in \eqref{out_dist} we can observe the expected behavior in the limits
$\lambda,r\rightarrow 0$ and $\lambda,r\rightarrow\infty$.
So the two-parameter distribution \eqref{out_dist} includes the Poisson
distribution and the geometric distribution as special cases.
\figref{fig:deg_dist} shows an example of the resulting distributions for $r=1$ and
$\lambda=1/5$.

As a function of $r$, the mean degree increases approximately as $\pi  r^2$ for small $r$ (this is the
region where the degree is power-limited)
and has a cap at $1/\lambda$ (due to the secrecy condition). Hence there exists
a power-limited and a secrecy-limited regime, and the inflection
point of $\E N(r)$, which is $r_T=(2\pi\lambda)^{-1/2}$ is a suitable boundary.
This is illustrated in \figref{fig:mean_degree}. Generally, the curve
$2\pi r^2=1/\lambda$ separates the two regimes.
Note that in the {\em po}wer-limited regime, the distribution is close to {\em Po}isson,
whereas in the secrecy-limited regime, it is closer to geometric.
Using the maximum slope $s$ of $\E\Nout(r)$,
a simple piecewise linear upper bound on the mean degree is
\begin{equation}
  \E\Nout(r) < \min\{s r, \frac1\lambda\}\,,\quad s\triangleq \sqrt\frac{2\pi}{e\lambda}\,.
\end{equation}
This bound is reasonably tight for $\lambda$ not too small.

As a function of $\lambda$, the mean degree is monotonically decreasing from $\pi r^2$ to 0, upper bounded by $1/\lambda$.

The transmission range (power) needed to get within $\epsilon$ of the 
maximum mean out-degree (for $r=\infty$) is
\begin{equation}
  r_\epsilon=\sqrt\frac{-\log\epsilon}{\lambda\pi}
\end{equation}
For example, $r_{0.01}=1.21/\sqrt\lambda$ achieves a mean out-degree of $0.99/\lambda$.

Next we establish bounds on the node degree distribution in the basic
graph $G_{\lambda,\infty}$.
Let $R$ be the distance of the nearest bad guy. If the second-nearest
bad guy is at distance at least $2R$, which happens with probability
$\exp(-\lambda\pi 3R^2)$, then bidirectional secure communication
is possible to any good guy in the area $a\pi R^2$ where $a=2/3+\sqrt 3/(4\pi)\approx 0.80$ (circle minus
a segment of height $R/2$). 
As a lower bound, we consider the circle of radius $R/2$. For sure bidirectional
communication is possible to any node within that distance. (This bound would
be tight if there were many more bad guys, all at the same distance $R$.) So we have
\begin{equation*}
    \sum_{k=0}^n \E_R\left[\exp(-A)\frac{A^k}{k!}\right]  <\P[N\leqslant n]<
    \sum_{k=0}^n \E_R\left[\exp(-B) \frac{B^k}{k!}\right]
\end{equation*}
where $A=a\pi R^2$ and $B=b\pi R^2$ with $b=1/4$.
The bounds are geometric:
\begin{equation}
  1-\left(\frac{a}{a+\lambda}\right)^{n+1} <\P[N\leqslant n]< 1-\left(\frac{b}{b+\lambda}\right)^{n+1}\,
\end{equation}
Since $\E N=\sum\P[N>n]$, the bounds $a/\lambda > \E N > b/\lambda$
for the mean degree follow. From \eqref{mean_deg_basic} we already know that
$\E N=1/(\lambda c)\approx 0.62/\lambda$.

Lastly, in this subsection, we consider the enhanced graph.
\begin{proposition}
 The mean degree $\E N'$ in the enhanced graph $G'_{\lambda,r}$ is
 \begin{equation}
   \E N'=\frac 2\lambda(1-\exp(-\lambda\pi r^2))-\frac{1}{c\lambda}(1-\exp(-c\lambda\pi r^2))\,.
   \label{enprime}
 \end{equation}
\end{proposition}
\begin{IEEEproof}
 This follows by combining \eqref{degree_relation}, \eqref{mean_deg_dir}, and 
   \eqref{mean_deg_basic}. 
\end{IEEEproof}

\subsection{Secrecy ratios}
Using the mean degree established in \eqref{mean_deg_basic}
we obtain
\begin{equation}
  \eta(\lambda,r)= \frac{1-\exp(-c\lambda\pi r^2)}{c\lambda\pi r^2}\,.
  \end{equation}
$\eta(\lambda,r)$ is decreasing in both $r$ and $\lambda$.
$\eta'(\lambda,r)$ follows from \eqref{enprime}.
Of interest are also the relative edge densities of the enhanced and basic graphs:
\begin{fact}
At least a fraction $3\pi/(5\pi+3\sqrt 3)\approx 0.45$ of the edges in the enhanced
graph $G'_{\lambda,r}$
are present in the basic graph $G_{\lambda,r}$.
\end{fact}
The ratio $\E N/\E N'$ is $1$ for small $\lambda r^2$ and reaches its minimum
as $\lambda r^2\rightarrow\infty$, where it is $(2c-1)^{-1}$ with $c$ as in
\eqref{mean_deg_basic}. The consequence is that in some graphs,
more than 50\% of the links can only be used securely in one direction
(unless one-time pads are used).

\subsection{Edge lengths}
We consider the distribution of the length of the edges in $\vec G_{\lambda,\infty}$.
For each node, its nearest bad guy determines the maximum length of an out-edge.
So we intuitively expect the edge length distribution to approximately inherit the
distribution of the distance to the nearest bad guy. Simulation studies reveal that
indeed the edge length distribution is very close to 
Rayleigh with mean $1/(2\sqrt\lambda)$, with only very slightly higher probabilities
for longer edges---which is expected since nodes whose nearest bad guy is
far will have many long edges on average and thus skew the distribution.

In the power-limited regime, with $r$ finite and $\lambda\rightarrow 0$, the edge length
pdf converges to the usual $2x/r^2$, $0\leqslant x\leqslant r$.

\subsection{Oriented percolation of $\vec G_{\lambda,r}$}

We are studying oriented out-percolation in $\vec G_{\lambda,r}$, \ie,
the critical region in the $(\lambda,r)$-plane for which there is a positive
probability that the out-component containing the origin has infinite size.
\begin{fact}
 $\lambda_c(r)$ is monotonically increasing for $r>\rg$, and we have
\begin{equation}
  0<  \lim_{r\rightarrow\infty}\lambda_c(r)  < 1\,. 
  \label{limr}
 \end{equation}
 In other words, 
  there exists a $\lambda_\infty<1$ such that for $\lambda>\lambda_\infty$, 
  $G'_{\lambda,r}$ does not percolate for any $r$.
\end{fact}
This follows from the facts that for fixed $r$ the mean degree is continuously
decreasing to $0$ as a function of $\lambda$, and for $\lambda<1$, the mean degree is smaller
than 1 even for $r=\infty$, so percolation is impossible.
We will use $\lambda_\infty$ to denote the limit in \eqref{limr}.
For intensities smaller than that, we define
\begin{equation}
  r_c(\lambda)\triangleq \sup \{r\,:\: \theta(\lambda,r)=0 \}\,,\quad\lambda\leqslant \lambda_\infty\,.
\end{equation}

From the monotonic decrease of the mean degree in $\lambda$ follows:
\begin{fact}
The percolation radius $r_c(\lambda)$ is monotonically increasing with $\lambda$
and has a vertical asymptote at $\lambda_\infty$.
\end{fact}

\begin{conjecture}
\label{conj:convex}
 $r_c(\lambda)$ is convex (and, consequently, $\lambda_c(r)$ is concave):
 \begin{equation}
   \frac{\d^2 r_c(\lambda)}{\d\lambda^2} > 0\,\quad
   \forall\, 0\leqslant \lambda < \lambda_\infty
 \end{equation}
 It follows that
 \begin{equation}
   r_c(\lambda) \geqslant \rg+c\lambda\,,\quad\text{where }
      c\triangleq \frac{\d r_c(\lambda)}{\d\lambda}\Big|_{\lambda=0}\,. 
 \end{equation}
\end{conjecture}
Simulation results show that $\lambda_\infty\approx 0.1499$ with a standard
deviation of $5.8\cdot 10^{-4}$ over 200 runs. 

\begin{figure}[h]
\centerline{\epsfig{file=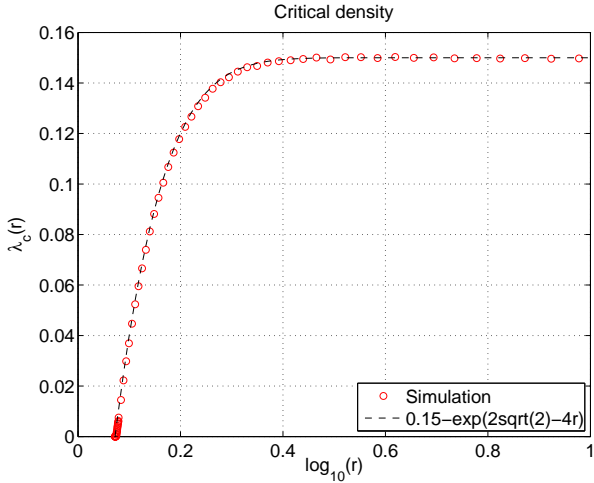,width=8cm}}
\caption{The simulated critical density $\lambda_c(r)$ vs.~$\log_{10}(r)$ together with a
  simple exponential approximation. Each simulated point is the average of 30 runs.}
  \label{fig:crit_density}
\end{figure}

Since $\lambda_c(r)$ is concave and converges to a finite $\lambda_\infty$,
we may conjecture that it can
be well approximated by a function of the form:
\begin{equation}
  \lambda_c(r)\approx \lambda_\infty-\exp(a-br)\,,\quad r>\rg\,,
\end{equation}
where $a$ and $b$ are related through $a=\log\lambda_\infty+b\rg$.
Indeed simulation results (see \figref{fig:crit_density})
reveal that for $b=4$, $\lambda_\infty=0.1499$, and
$a=2\sqrt 2$, we obtain an excellent approximation. Similarly,
\begin{equation}
 r_c(\lambda)\approx \frac a b-\frac 1 b \log(\lambda_\infty-\lambda)\,,\quad
  \lambda<\lambda_\infty\,.
 \label{rc_approx}
\end{equation}

It follows that the constant $c$ in Conj.~\ref{conj:convex} is
$c=(b\lambda_\infty)^{-1}=5/3$, and the slope of
$\lambda_c(r)$ at $r=\rg^+$ is $3/5$.

For the {\em critical} graph $\vec G_{\lambda,r_c(\lambda)}$,
it turns out that both $\P[N=0]$ and $\E N$ are increasing with $\lambda$.

A good empirically derived approximation is
\begin{equation}
  \P[\Nout_c=0]\approx \frac{1}{80}+\frac45\lambda\,,\quad\lambda<\lambda_\infty\,.
\end{equation}
For the mean out-degree we have from \eqref{mean_deg_dir} and \eqref{rc_approx}
\begin{equation} 
  \E \Nout_c(\lambda) > \pi \rg^2+\frac{11}4 \lambda\,.
\end{equation}
$\E \Nout_c(\lambda)$ is convex and reaches $1/\lambda_\infty$ at
$\lambda=\lambda_\infty$ per Prop.~\ref{prop:degree}.

So percolation on the secrecy graph requires a higher mean degree
than Gilbert's disk graph. Since the disk graph was shown to require
the highest mean degree among all germ-grain random geometric graphs
\cite{net:Balister04isit,net:Franceschetti05,BBW:ann},
we have established that:
\begin{fact}
 The secrecy graph is not equivalent to any germ-grain random geometric graph.
\end{fact}

\section{Concluding Remarks}
We have introduced a new class of random geometric graphs that captures
  the condition for secure communications in ad hoc networks.
For the lattice-based models, there exist direct analogies to bond and
     site percolation problems.
In Poisson-based networks, we have derived the mean node degrees and,
in some cases, the distribution. As a byproduct, a two-parameter distribution was found that
includes the Poisson and the geometric distribution as special cases. Based
on the mean degree, we defined power- and secrecy-limited regions in the
$(\lambda,r)$-plane.
  The percolation region $\{(r,\lambda)\,:\: \theta(r,\lambda)>0\}$ was
   bounded and numerically determined. In conclusion,
the presence of eavesdroppers is rather harmful in the random case.
  A (relative) density of $0.15$ is sufficient to make percolation impossible.
Many interesting problems remain open; we hope that this initial study sparks
further investigations.
  
\section*{Acknowledgments}
The author wishes to thank Aylin Yener for the discussions leading
to the problem studied in this paper. The support of NSF
(grants CNS 04-47869, DMS 505624, and CCF 728763),
and the DARPA/IPTO IT-MANET program
(grant W911NF-07-1-0028) is gratefully acknowledged.

\bibliographystyle{IEEEtr} 
\bibliography{header,comm,bela,net}

\end{document}